# Risk analysis for long term disposal of radioactive nuclear waste

Sabikun Nahar Sadia[1], Md. Iqbal Hosan[1], Md. Jafor Dewan[1], Saikat Ahmed[2]
[1]Department of Nuclear Engineering, University of Dhaka, Dhaka-1000, Bangladesh
[2] Rooppur NPP, Bangladesh Atomic Energy Commission, Dhaka-1000, Bangladesh
Corresponding author's Email: iqbalhosan@du.ac.bd

**Abstract:** In this study, the technical risk analysis that was conducted using fault trees. It was found that the probability for the radioactive material to reach through the various paths designed is approximately $2.508 \times 10^{-12}$. This means that it has no immediate impact on the environment even if the barrier system fails to work. With respect to the scenarios in Bangladesh, there is still no concrete proposal about a permanent disposal method, so this analysis is based on the propositions and other historical data. When actual data is produced, this risk analysis technique can then be employed to calculate risks. Also, the nature of the geological structure will not remain constant throughout. Changes in the geological background might affect the working of the barrier. Hence, continuous monitoring and update of the numbers are essential.
**Key points:** Radioactive waste, disposal of waste, Risk analysis technique, FMDA, Fault tree method

**Introduction**

The present system of electricity production is not only inadequate but also the harm to the nature is more from this system. Burning of fossil fuel produces a heavy amount of greenhouse gases which damage the equilibrium of nature. These gases cause the temperature of the earth to increase which leads to the melting of polar ice. Thus the sea level rises and more and more land goes under water [1]. So it is time to go beyond the traditional lines of production of electric energy and bring about a significant change in its production. To keep pace with the increasing demand it needs a source that can produce much more electricity than the present production. Such a breakthrough in electricity production can only be achieved through the introduction of nuclear energy. Nuclear energy is produced by fission reaction of radioactive metals in a nuclear reactor [2]. The amount of energy produced from nuclear reaction is gigantic compared to the energy produced from other primary resources [3].

Nuclear power plants also have a higher rate of efficiency compared to other primary energy based power plants. On the other hand, nuclear energy is considered as a clean and environmental friendly source of energy, if compared to other resources like oil, coal or gas, the consequences of nuclear incidents may have devastating effects on human health and the environment. Not only nuclear incidents are such that may have adverse effects on human health and the environment, but also detrimental management of radioactive waste and spent fuel, which are left after the use of nuclear materials, may produce huge adverse impacts on human health and the environment [4].

Therefore, the controversy around nuclear energy raises a lot of issues especially those regarding the management of radioactive waste and spent fuel. When one compares radioactive waste and spent fuel with other types of waste, it can be concluded that other types of waste are much more represented worldwide. Therefore, it appears that the percentage of radioactive waste and spent fuel is rather small if it is compared with the percentage amount of other types of waste. The main difference is that this small percentage may create worldwide adverse consequences on human health and the environment. Managing radioactive waste and spent fuel requires a lot of scientific effort and number of legal instruments in order that safe and sound management may be provided and humans and the environment properly protected [5].

Various types of waste produces from the nuclear power plant are needed to handle with high level of safety [6]. The risks associated with the management of waste are analyzed in this study by fault tree analysis technique [7, 8]. Fault tree analysis uses simple logical relationships like AND, OR, XOR etc. to relate events that interact to produce other undesirable events. This allows us to build a methodical structure that represents the system as a whole. In order to construct a fault tree, it is mandatory to understand the functioning of the system under analysis. A flow diagram or a functional logic diagram can be used for this purpose. Another method used frequently to understand failure modes is Failure Modes and Effects Analysis (FMEA) [9]. When constructing a fault tree, the first step to it is selecting the top failure event. The events following it will be considered according to the effect they place on the top event. The next step is to find contributing events that will lead to the occurrence of the top event. There are at least four possibilities that exist (these are applicable to any kind of fault tree construction): (i) Failure of the device to receive an input signal, (ii) Failure of the device itself to operate, (iii) Human error, and (iv) Failure due to external events [10].

The purpose of FMEA analysis is to provide a systematic analysis method to identify potential failure modes of systems, components and/or assemblies. The analysis provides input to the design team on how to mitigate the risk of potential failures to an acceptable level. Failures should be prioritized according to how serious their consequences are, how frequently they occur and how easily they can be detected. Action to eliminate or reduce failures should begin with those with the highest priority.

**Analysis and Results**

In this study some engineering and physical barriers which prevent escape of radio nuclides into the environment were identified. In the case of processing plants, barriers include: piping, valves, pumps, physical or radiological protection, etc. In the case of the disposal facility, the multi-barrier protective shield consists of three systems: (i) The matrix (container), which contains the RAW (Radioactive Waste), (ii) The repository as an engineering and construction structure, and (iii) The geological environment in which the repository is located.

To formalize the numerical calculations, the functioning of the individual protection system is described by the binary relation of 'failure' to 'work'. Decomposition of the protective shield to the individual systems is an essential part of the safety analysis. An emergency event which may be an incident, accident, occurrence, or situation, etc., is defined by a sequence of failures of the systems and components leading to adverse effects such as a loss of control over the ionizing radiation sources or uncontrolled escape of energy and matter from a dangerous object into the environment. This sequence of events is called a scenario. If a complex of the protective shield is divided into $N$ systems, there are only $M = 2^N$ scenarios. In the case of near-surface disposal facility, $N = 3$ and $M = 8$. Functioning of the complex is displayed graphically in Figure. 1. Such a diagram is called an event tree.

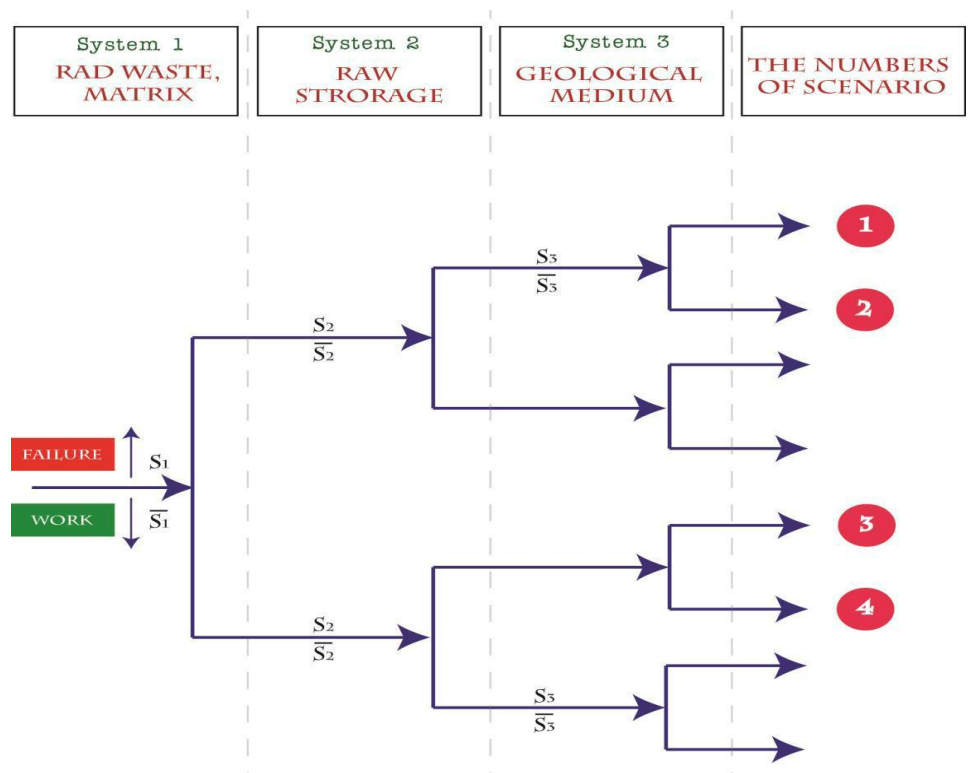

Figure. 1 Emergent event trees for activity of disposal complex.

In this diagram, the systems are presented in the form of columns. The rows represent the state of these systems. Armed with the event tree, it is easy to calculate the probability of each scenario from the general formula:

$$P_m = \prod_{n=1}^{N} S_n^{(1-q_{nm})} \bar{S}_n^{q_{nm}}$$

Where, $S_n = 1, 2, \ldots, N$, is the failure probability of the $n$-th system, and $\bar{S}_n = 1 - S_n$ is the probability of the work state of this system; $m = 1, 2, \ldots, M$ is the scenario number; $q_{nm}$ is the state indicator of the system; $q_{nm} = 0$ if in scenario $m$ the system $n$ is in a failure state, and $q_{nm} = 1$ if the system is in a working state.

The probabilities $S_n$ are calculated by analysis of the failure trees for each system. The failure tree is a logical connection between the system elements, connected by symbols 'OR', 'AND', corresponding to addition or multiplication of the random failure events. The symbol 'OR' links together the group of elements, failure of at least one of which leads to failure of the entire group. The failure probability of such a group is calculated from:

$$S_{j_{or}} = 1 - \prod_{j=1}^{j_{or}} (1 - E_j)$$

Where $E_j$ is the failure probability of the $j$-th element from the group; $j_{or}$ is the number of elements in the group. The symbol 'AND' combines the group of elements, only the joint failure of which leads to failure of the whole group. The failure probability of such a group is calculated from:

$$S_{j_{and}} = \prod_{j=1}^{j_{and}} E_j$$

Where $j_{and}$ is the number of elements in the group.

The failure probability of elements is calculated from:

$$E_j = 1 - e^{(-\lambda_j t)}$$

Where $\lambda j$ is the failure rate of a given element; and $t$ is the time from start of observation or operation of the disposal facility.

Next step is to analysis of alarm events associated with possible escape of radionuclides into the environment from the complex for disposal of RAW. For this consider only the four scenarios of the alarm events that have physical meaning

- Scenario 1. Failure of all systems that make up the disposal complex.
- Scenario 2. Joint failure of the matrices and the disposal facility.
- Scenario 3. Only failure of system 2 (the repository). Physical representation of this scenario is to destroy the structural elements of the repository with probable leakage of radionuclides beyond.

• Scenario 4. Only joint failure of systems 2 (the repository) and 3 (the geological environment). Physical representation of this scenario is the escape of radionuclides from the disposal facility and their migration into the geological environment.

As the scenarios are interdependent, the sum of probabilities of all scenarios is 1. Therefore the probability of accident, $P_{ac} = P1 + P2 + P3 + P4$ and

The probability of the work, $P_{work} = 1 - P_{ac}$.

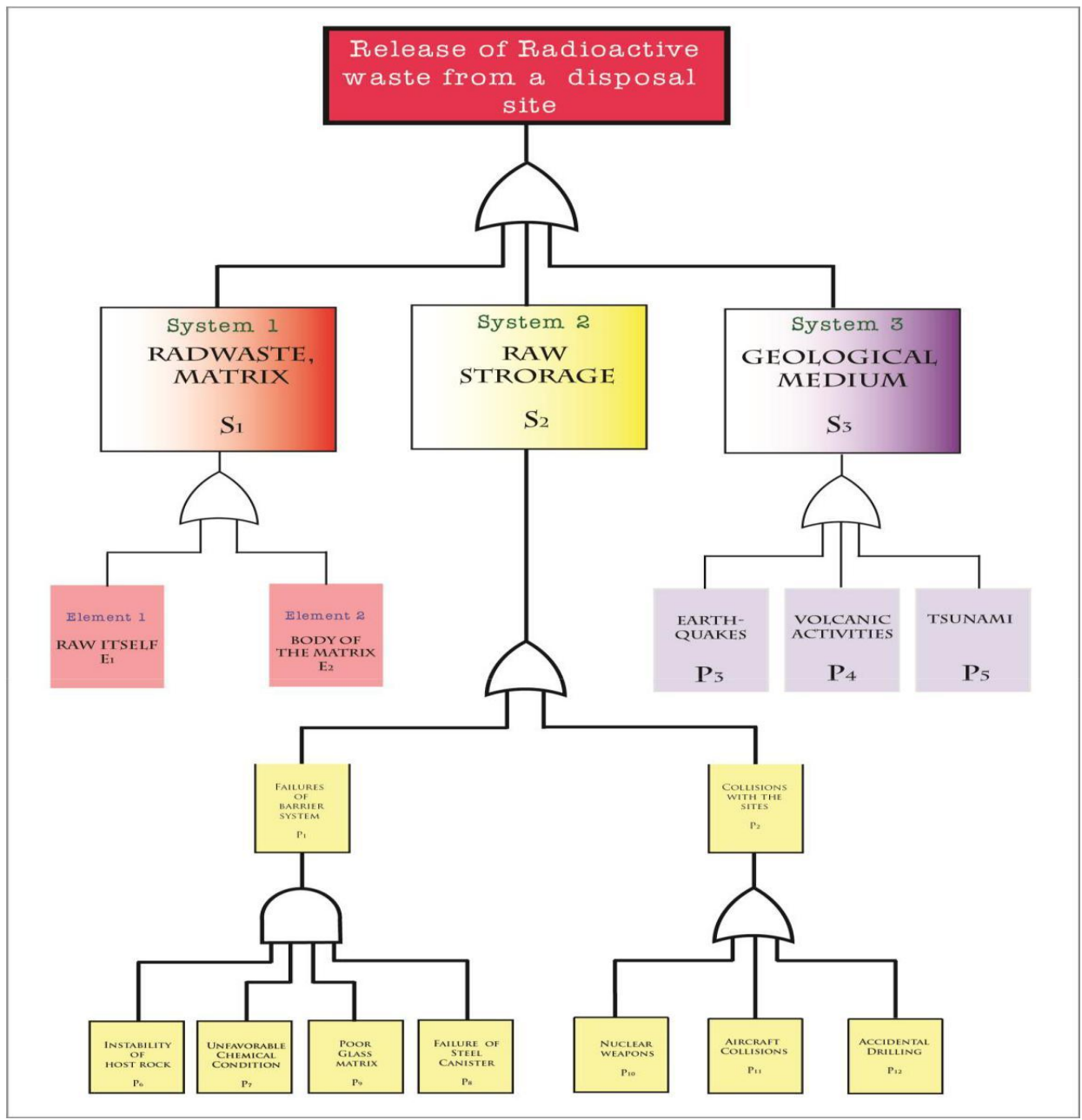

Figure 2: Fault tree model for failure during long-term disposal

$$P = S1 + S2 + S3$$

In the model representation, system 1 consists of two elements. Element 1 is the RAW itself, contained in a matrix or container. Element 2 is the body of the matrix. System 1 failure occurs when a failure occurs in element 1 or element 2 or both. Hence, these two elements are working on an 'OR' scheme. The relevant failure tree is shown in Fig. 3. Physical representation of element 1 failure is the radionuclide escape from the matrix body as a result of diffusion and leaching. Physical representation of the element 2 failure is the matrix degradation during its aging, corrosion, and cracking.

Although the physical processes of failure of these two elements are interrelated, from a model point of view it is convenient to present them as independent.

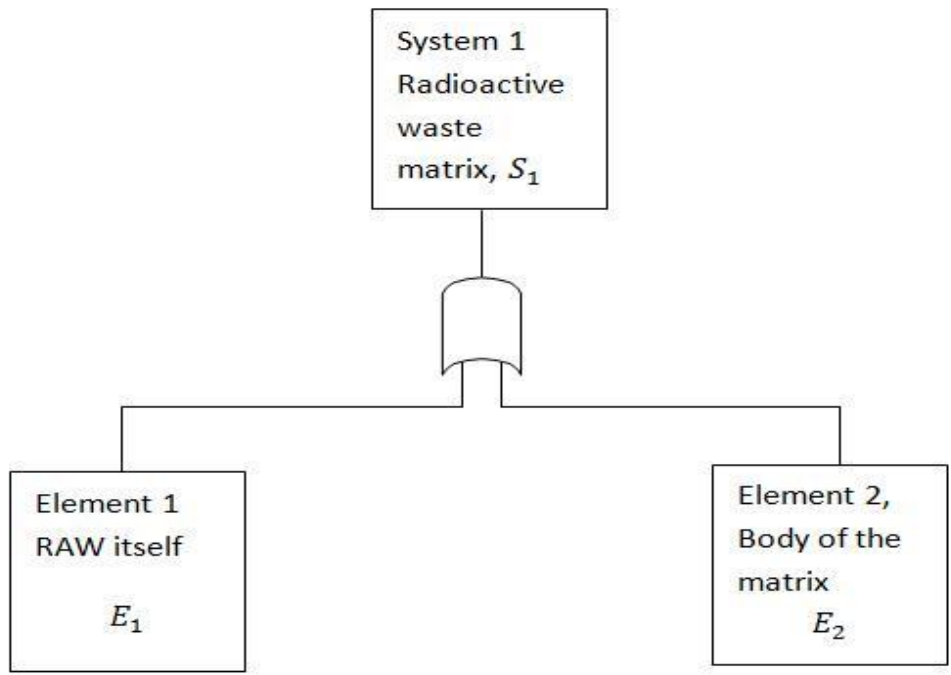

Figure 3. Failure tree of matrix with radioactive waste

From Eq .3 we calculate the failure probability of elements 1 and 2.
Then the failure probability of system 1 can be calculated from Eq. 2

$$S_1 = 1 - (1-E_1)(1 - E_2)$$

We now discuss the physical barrier in the RAW repository (system 2). From the system analysis standpoint, this system can be regarded as consisting of the following elements: covering slabs, walls, a bottom, waterproofing.

In turn, these elements may be composed of elementary units: concrete slabs, cement joints, beams, etc. The failure tree of system 2 is shown in Fig. 4. It should be noted that, depending on whether the composition of a structural element contains slabs or joints, they may have different performance parameters.

The system 2, as seen in the below figure, there are two possible pathways of which one is an engineered effect. The first level of sub-ordinate events (Failure of barrier system, Collisions with the site) are all related with an OR gate to the top event. Any kind of collision that can have a huge impact on the repository site is an undesirable event and three paths have been identified for this; 1. Activities caused by nuclear weapons; war etc 2. Aircraft collisions and 3. Accidental drilling in the disposal site. Of the three, accidental drilling has a higher probability of occurrence than the other two events. The probability calculations for each of the events will be shown in the following sections.

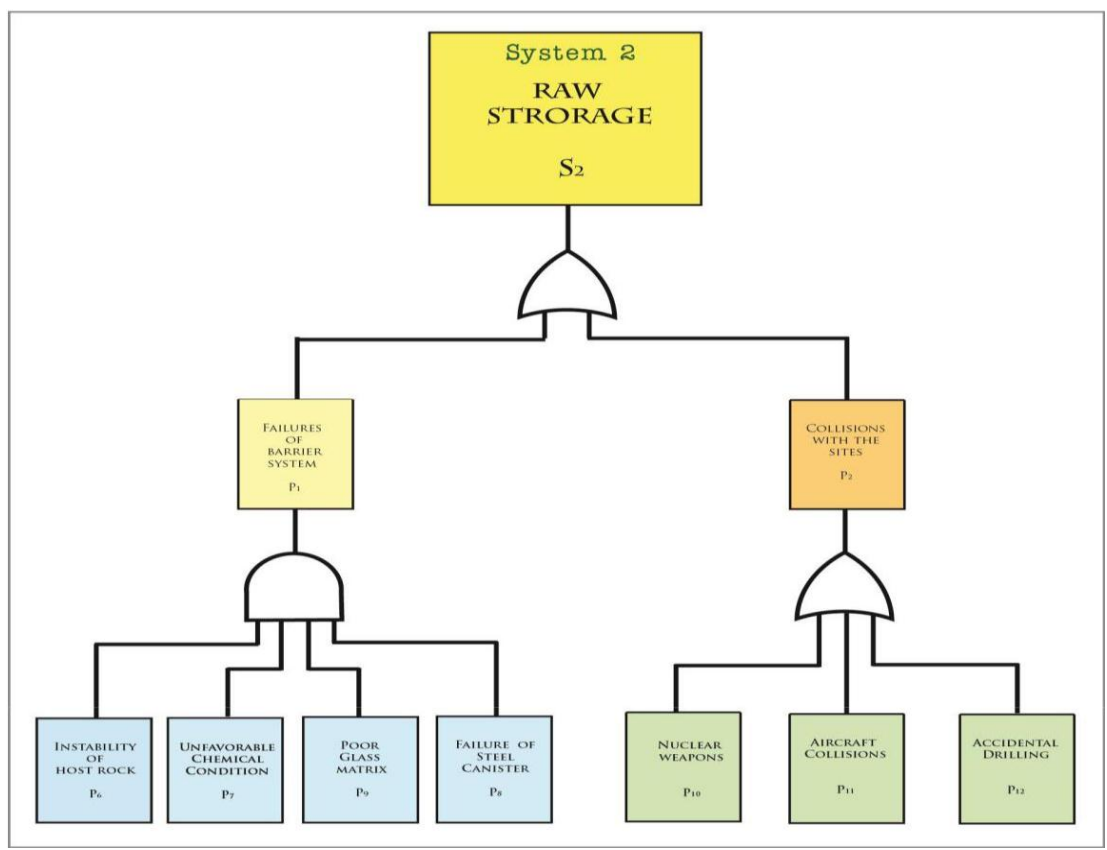

Figure: 4 Failure tree for raw storage

$$S_2 = P_1 + P_2$$

$$P_1 = P_6 * P_7 * P_8 * P_9$$

$$P_2 = P_{10} + P_{11} + P_{12}$$

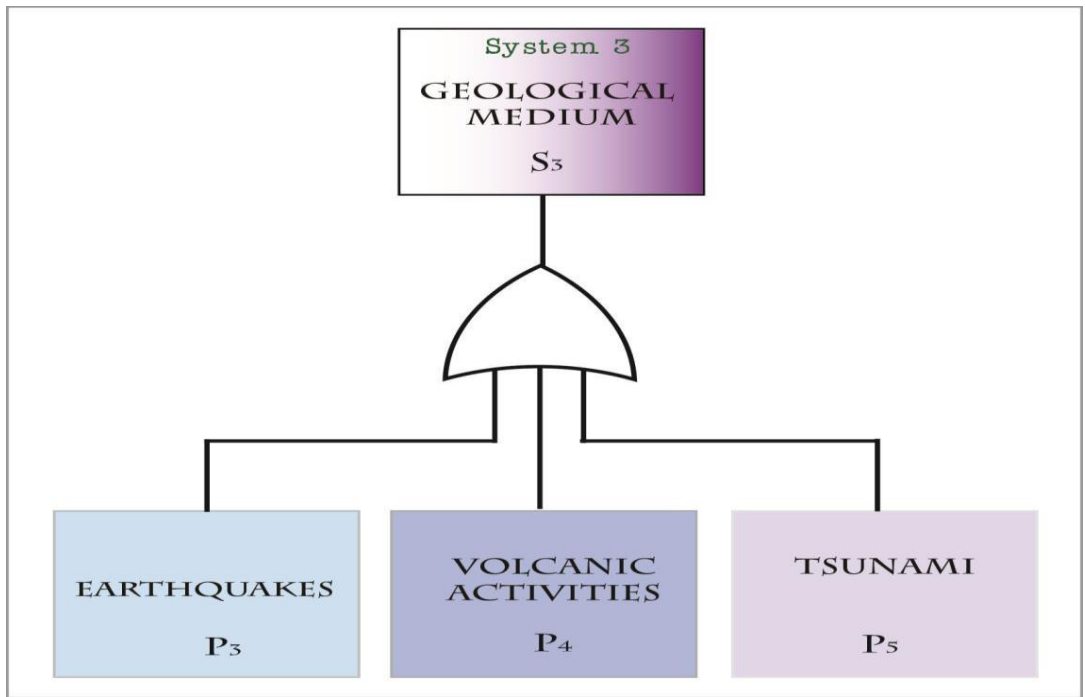

Figure 5: Failure tree of Geological Medium

$$S_3 = P_3 + P_4 + P_5$$

External events such as volcanic activities or earthquakes or tsunami depend on the geography of the location. Probability estimation of such kind of seismic activities cannot be predicted correctly. Timescale for such predictions is very hard to set.

Let's consider the failure of the barrier system. The waste from the nuclear power plant is permanently disposed in a repository which is a multiple barrier system consisting of a combination of both natural and engineered barrier. After the waste package is placed and the tunnel is filled, the engineered barrier will begin to degrade. Gradual degradation of the barriers caused by the increased level of inflowing groundwater is by far, the most likely failure mechanism. The Bentonite buffer will begin to swell and adsorb the water that reinvaded into the zones of the host rock. But because of the low hydraulic conductivity of the backfill, this resaturation process will take a long time to completely have some effect on the barrier system. Corrosion will occur when this wetting comes in contact with the steel canister. The rate of corrosion is affected by the build of oxyhydroxides. Once the canister is corroded, the fluid will come into contact with the glass matrix. Once the fluid is out of the matrix, it is transported to the surface of the environment and thus comes in contact with the environment. But the rate at which the transportation occurs depends entirely on the rate of diffusion of the Bentonite pores, the magnitude of different retardation process that will occur and on the level of water flow into the host rock

Following the laws Boolean expression, the below equations are obtained for each of the levels.

The probability for failure during long-term disposal of waste is
$$P = S_1 + S_2 + S_3$$
Substituting in the expression for top event, we get;
$$P = 1 - (1-E_1)(1 - E_2) + (P_6 * P_7 * P_8 * P_9) + (P_{10} + P_{11} + P_{12}) + (P_3 + P_4 + P_5)$$

Using FMEA we find these unknown probabilities:

$\lambda 1 = 9.1 \times 10^{-14}$ 1/year

$\lambda 2 = 1.12 \times 10^{-12}$ 1/year

$E1 = 9.1 \times 10^{-14}$

$E2 = 1.12 \times 10^{-12}$

$S1 = 1.211 \times 10^{-12}$

$P6 = 2.60 \times 10^{-9}$

$P7 = 1.551 \times 10^{-9}$

$P8 = 1.111 \times 10^{-12}$

$P9 = 1.6 \times 10^{-11}$

$P1 = P6 \times P7 \times P8 \times P9 = 7.168 \times 10^{-41}$

$P10 = 1.58 \times 10^{-14}$

$P11 = 1.253 \times 10^{-15}$

$P12 = 1.091 \times 10^{-12}$

$P2 = P10 + P11 + P12 = 1.108 \times 10^{-12}$

$S2 = P1 + P2 = 1.108 \times 10^{-12}$

$P3 = 1.683 \times 10^{-13}$

$P4 = 4.091 \times 10^{-18}$

$P5 = 2.091 \times 10^{-14}$

$S3 = P3 + P4 + P5 = 1.892 \times 10^{-13}$

$P = 1 - (1-E_1)(1 - E_2) + (P_6 * P_7 * P_8 * P_9) + (P_{10} + P_{11} + P_{12}) + (P_3 + P_4 + P_5) = 2.508 \times 10^{-12}$

Substituting the probabilities of each failure event in the above equation, the probability for failure during long-term disposal of waste is 2.508 x 10 $^{-12.}$ The FMEA method that is employed to predict the unknown probabilities is shown in Appendix A.

**Conclusion**

Nevertheless, geological disposal is a safe and viable option and the engineered barrier system will prevent the radioactive materials from reaching the environment. Safe disposal is definitely feasible with current technologies. Waste is a byproduct of all industries and nuclear power is responsible for some types of waste which are particularly unattractive. But such wastes are produced only in low quantities when compared to the end product value which is electricity. Hence repositories involving high-level barrier system design, can be used to provide a high degree of safety. The only issue now is to convince the public that such kind of technology is safe to use. An important consideration for any new nuclear program is an understanding of the mass's viewpoint, as in many countries this can influence the direction of future energy markets.

11. http://asq.org/learn-about-quality/process-analysis-tools/overview/fmea.html (accessed 15th January 2018)

**Appendix A: FMEA Calculation**

The following FMEA calculation is done to determine failure rates for those events whose probabilities were not known. In practices, such calculations are done in groups but because this is an individual presentation, the below calculations are made by just one person. [11]

| Steps (for 1 year) | Failure mode | Failure causes | Failure effects | Likelihood of occurrence (1-10) | Likelihood of detection (1-10) | Severity (1-10) | Risk profile number | Action |
|---|---|---|---|---|---|---|---|---|
| unfavorable chemical conditions | will lead to the failure of the barrier system | changes in the chemical conditions of the radioactive material stored | will result in the release of radioactive material into the environment | 3 | 3 | 7 | 63 | ensure favorable chemical conditions of the material |
| Increased oxygen content | will cause corrosion of steel container | increase in oxygen level in the repository site | corrosion of steel thereby leading to leakage of stored radioactive waste | 3 | 2 | 8 | 48 | provide corrosion resisters steel container |

| inflowing ground water increase | will transport radioactive waste to the surface | changes in geographical structure, leading to cracks | will lead to explore of radioactive material | 3 | 2 | 8 | 48 | careful location on of site |

**Appendix B: RPN**

One common approach in FMEA calculates a Risk Priority Number (RPN). Each failure (mode) has an assigned severity, probability, and detectability values. This common approach uses the following qualitative scale for ranking.

- The severity score (S) is an integer between 1 and 10, where the most severe is 10.

- The probability score (P) is an integer between 1 and 10, where the highest probability is 10.

- The detectability score (D) is an integer between 1 and 10 where most difficult to detect is 10.

The RPN is the product of the three ranks. For example, with S = 5, P = 3, D = 6, the RPN is 90 = (5)(3)(6).

The RPN is not a measure of risk, but of risk priority. You would apply your limited resources to the most important problems. The RPN gives you a model to allocate these resources. Higher numbers are higher priority, so you should work on an RPN of 900, before you put resources on an RPN of 30.

One interesting issue, not well understood, is that some numbers cannot be an RPN. Many people have the mistaken belief that any number from 1 to 1,000 can be an RPN. Consider 17. It cannot be an RPN because it is a prime number larger than 10. You cannot multiply 3 numbers from 1 to 10 and get 17.

In contrast, many numbers can occur in multiple ways, the highly composite numbers. Consider 120. There are 24 different ways it can become an RPN. Two examples are (S)(P)(D) = (2)(6)(10) = (8)(5)(3). In the first case the severity is near the bottom of the scale, 2, while in the second case, the severity is near the top of the scale, 8. Also, note that there are long stretches of numbers that can't be an RPN. For example, no number from 901 to 999 is an RPN. There are only 120 possible RPN values.